\documentclass[10pt]{amsart}
\usepackage{amsfonts,amssymb,amsmath,amsthm}
\usepackage{url}
\usepackage{enumerate}
\usepackage{xcolor}
\newtheorem{thrm}{Theorem}[section]
\newtheorem{lem}[thrm]{Lemma}
\newtheorem{prop}[thrm]{Proposition}
\newtheorem{cor}[thrm]{Corollary}
\newtheorem{definition}[thrm]{Definition}

\numberwithin{equation}{section}

\newcommand{\F}{\mathbb{F}}

\newcommand{\CC}{\mathcal C}

\newcommand{\vr}{\vspace*{0.5cm}}

\title{Good codes from twisted group algebras}

\author{Samir Assuena } 
\address{Centro Universit\'ario da FEI,
Av. Humberto de Alencar Castelo Branco 3972, S\~ao Bernardo do Campo-SP, Brazil, CEP: 09850-901.} 
\address{Instituto de Matem\'atica e Estat\'\i stica da USP (IME-USP), Rua do Mat\~ao 1010, S\~ao Paulo-SP, Brazil, CEP: 09210-580.}  \email{samir.assuena@fei.edu.br and samir.assuena@alumni.usp.com.br}


\begin{document}

\begin{abstract}
In this paper, we shall give an explicit proof that constacyclic codes over finite commutative rings can be realized as ideals in some twisted group rings. Also, we shall study isometries between those codes and, finally, we shall study $k$-Galois LCD constacyclic codes over finite fields. In particular, we shall characterize constacyclic LCD codes with respect to Euclidean inner product in terms of its idempotent generators and the classical involution using the twisted group algebras structures and find some good LCD codes.
\end{abstract} 
\maketitle

\section{Introduction}

Linear codes with complementary duals (abbreviated LCD) are linear codes whose intersection with their dual is trivial. When they are binary, they play an important role in armoring implementations against side-channel attacks and fault injection attacks. 

Linear complementary dual codes have importance in data storage, communications systems and security too.

These codes have been studied for improving the security of information on sensitive devices against side-channel attacks (SCA) and fault non-invasive attacks, see \cite{CC}, and have found use in data storage and communications systems.

Carlet and Guilley, in \cite{CG}, also investigated the application of binary LCD codes against side-channel attacks (SCA) and fault tolerant injection attacks (FIA). Also, in \cite{TD}, the authors constructed explicity LCD codes and have explicit efficient encoding and decoding algorithms .

In \cite{FZ}, Fan and Zhang, introduced the concept of \textit{k-Galois form}, which is a generalization of Euclidean and Hermitian inner products and Liu, Fan and Liu, in \cite{LFL}, studied $k$-Galois LCD codes. 

So, this paper is devoted to study constacyclic codes in terms of twisted group rings of cyclic groups and to classify the $k$-Galois LCD constacyclic codes over finite fields in terms of  idempotents. Using that approach, we can find some good code from twisted group ring.

Let $R$ be a finite commutative ring, $\CC$ be a linear code over $R^{n}$, that is, $\CC$ is a $R$-submodule of $R^{n}$ and let $\lambda$ be an element of ${\mathcal{U}}(R)$, the group of units of $R$. We say that $\CC$ is a \textit{$\lambda$-constacyclic code} if 

\begin{center}
$(c_{0}, c_{1}, \cdots, c_{n-1}) \in \CC \Longrightarrow (\lambda c_{n-1}, c_{0}, \cdots, c_{n-2}) \in \CC $
\end{center}

\noindent for all $(c_{0}, c_{1}, \cdots, c_{n-1}) \in \CC$.

When $\lambda=1$, we have so called \textit{cyclic codes} and, when $\lambda=-1$, we have \textit{negacyclic codes}. Thus, constacyclic codes are generalization of cyclic and negacyclic codes and they have been studied for many authors (\cite{BR}, \cite{CDLW}, \cite{D}). Also, constacyclic codes can be realized as ideals in polinomial factor ring $\displaystyle\frac{R[x]}{\left\langle x^{n}-\lambda\right\rangle}$.

Given $x=(x_{0}, x_{1}, \cdots, x_{n-1})$ and $y=(y_{0}, y_{1}, \cdots, y_{n-1})$ two elements of a linear code  $\CC$, the \textit{Hamming distance} between $x$ and $y$ is the number

\begin{center}
	$d_{H}(x,y)=|\{i, \,\, x_{i}\neq y_{i}, \,\, 0\leq i\leq n-1\}|$.
\end{center}

\noindent and the \textit{weight} of $x$ is 

\begin{center}
	$w_{H}(x)=d(x,0)=|\{i, \,\, x_{i}\neq 0, \,\, 0\leq i\leq n-1\}|$.
\end{center}

It is well-known that, for a linear code $\CC$, we have $d_{H}(x,y)=w_{H}(x-y)$, for all $x,y \in \CC$.

Let $G$ be a group and $A$ be an abelian group. A map

\begin{center}
$\alpha: G\times G \longrightarrow A$
\end{center}

\noindent is a 2-\textit{cocycle} if , for all $x,\,y$ and $z$ in $G$, we have

\begin{center}
$\alpha(x,y)\alpha(xy,z)=\alpha(y,z)\alpha(x,yz)$.
\end{center}

\noindent and a map $t:G\times G \longrightarrow A$ is a 2-\textit{coboundary} if there is a map $\delta:G \longrightarrow A $ such that

\begin{center}
$t(x,y)=\delta(x)\delta(y)\delta(xy)^{-1}$.
\end{center}

As usual, the set of all 2-cocycles will be denoted by $Z^{2}(G,A)$ and the set of all 2-coboundary will be denoted by $B^{2}(G,A)$. Finally, we say that a 2-cocycle $\alpha$ is \textit{normalized} if $\alpha(x,1)=\alpha(1,x)=\alpha(1,1)=1$, for all $x \in G$. Notice that, if $\alpha$is a 2-cocycle, we can replace $\alpha$ by $\alpha'$ given by 

\begin{center}
$\alpha'(x,y)=\displaystyle\frac{\alpha(x,y)}{\alpha(1,1)}$
\end{center}

\noindent which is a normalized 2-cocycle.  From now on, we assume that all 2-cocycle are normalized.

Let $R$ be a commutative ring and $G$ be a group. The \textit{twisted group ring $R^{\gamma}G$ of G over} $R$ is the associative ring with basis $\overline{G}=\{\overline{g}, \, g\in G\}$, which is a copy of $G$, and the multiplication is defined on the basis as

\begin{center}
$\overline{g}\cdot \overline{h}=\gamma(g,h)\overline{gh}$
\end{center}

\noindent where $\gamma(g,h)$ is an element of ${\mathcal{U}}(R)$, the group of units of $R$.

The mapping $\gamma:G\times G \longrightarrow {\mathcal{U}}(R)$ is called \textit{twisting} and there are many different possibilities for $R^{\gamma}G$ depending on the choice of the twisting. For instance, the group ring $RG$ of $G$ over $R$ is a twisted group ring with $\gamma(g,h)=1$. Furthermore, the associative condition on the multiplication implies that

\begin{center}
$\gamma(g,h)\gamma(gh,k)=\gamma(h,k)\gamma(g,hk)$
\end{center}

\noindent and, for this reason, $\gamma$ is a 2-cocycle. 

When $G=C_{n}=\left\langle g\right\rangle$, a cyclic group of order $n$ and $R=\F$, a field, we have the following well-known result. See for example, \cite{KAR4}, Theorem 3.1.

 \begin{lem}\label{lem1}
 	Let $C_n = \langle g \rangle$ be a cyclic group of order $n$ and $A$ be a finite $C_n$-module, i.e., $A$ is a finite abelian group with an action of $C_n$ in $A.$  Let $A^{C_n} = \{a \in A : a^{g^i} = a\hspace{.1cm} \text{for all $g^i\in C_n$}\}$. Also, define the \textit{norm map} $N: A \rightarrow A^{C_n}$ by $N(a) = \prod_{i=0}^{n-1} a^{g^i}$.
 	
 	Then, for every $\lambda \in A^{C_n},$ we have that $\gamma_\lambda:C_n\times C_n \rightarrow A$ defined by
 	\begin{equation*}
 		\gamma_\lambda(g^i, g^j) = \begin{cases}
 			1, \quad i+j<n\\
 			\lambda, \quad i+j \ge n
 		\end{cases}
 	\end{equation*}
 	is a 2-cocycle and $H^2(C_n,A) = \{ [\gamma_\lambda ] : \lambda \in A\} \cong A^{C_n} / Im(N).$
 \end{lem}

It is possible make a \textit{diagonal} change of basis by replacing each $\overline{g}$ by $\widetilde{g}=\delta(g)\overline{g}$ for some $\delta(g) \in {\mathcal{U}}(R)$ and, with this change of basis, $R^{\gamma}G$ is realized in a second way as a twisted group ring of $G$ over $R$ with twisting

\begin{center}
$\widetilde{\gamma}(g,h)=\delta(g)\delta(h)\delta(gh)^{-1}\gamma(g,h)$.
\end{center}

In this case, we say that $\gamma$ and $\widetilde{\gamma}$ are \textit{cohomologous}.

\begin{lem} \label{1.1} {\rm{\cite [Lemma 2.1]{PA}}}
The following relations hold in $R^{\gamma}G$

\begin{enumerate}
	\item [i.] $1=\gamma(1,1)^{-1}\overline{1}$
	\item [ii.] For all $g \in G$,

\begin{center}
$\overline{g}^{-1}=\gamma(g,g^{-1})^{-1}\gamma(1,1)^{-1}\overline{g^{-1}}=\gamma(g^{-1},g)^{-1}\gamma(1,1)^{-1}\overline{g^{-1}}$
\end{center}
	
\end{enumerate}

\end{lem}

Let $C_{n}=\left\langle g \mid g^{n}=1\right\rangle$ be a cyclic group of order $n$, $R$ be a commutative ring and $R^{\gamma}C_{n}$ the twisted group algebra with 

\begin{center}
		$\gamma_{\lambda}(g^{j},g^{k})= \left\{
		\begin{array}{lll}
			\lambda,    & {\rm{if}} \,\, j+k\geq n\\
			1, &  {\rm{if}} \,\, j+k < n
		\end{array}
		\right.$
\end{center}

\noindent where $\lambda$ is a unit element of $R$. Thus, $\overline{g}^{2}=\overline{g}\cdot\overline{g}=\gamma(g,g)\overline{g^{2}}$, so we can make a diagonal change of basis and replace $\overline{g^{k}}$ by $\overline{g}^{k}$, for all $k, \,\, 1\leq k\leq n$. Thus, there exists a unit element $a \in R$  such that $\overline{g}^{n}=a\cdot 1$ which implies that $R^{\gamma}C_{n}$ is a commutative ring.

\section{Constacyclic codes over finite commutative rings}

In this section, we shall study constacyclic codes over a finite commutative ring.

\begin{thrm} \label{1.2}
Let $R$ be a finite commutative ring , $C_{n}=\left\langle g \mid g^{n}=1\right\rangle$ a cyclic group of order $n$ and $\CC$ be a linear code over $R^{n}$. Consider the linear mapping $\varphi: R^{n} \longrightarrow R^{\gamma}C_{n}$ given by $\varphi(c_{0}, c_{1}, \cdots, c_{n-1})=c_{0}\overline{1}+c_{1}\overline{g}+\cdots+c_{n-1}\overline{g^{n-1}}$. Then, $\CC$ is a $\lambda$-constacyclic code if and only if $\varphi(\CC)$ is an ideal of $R^{\gamma}C_{n}$ where 

\begin{center}
$\gamma_{\lambda}(g^{j},g^{k})= \left\{
\begin{array}{lll}
\lambda,    & {\rm{if}} \,\, j+k\geq n\\
1, &  {\rm{if}} \,\, j+k < n.
\end{array}
\right.$
\end{center}
\end{thrm}

\begin{proof} 
Let $\CC$ be a linear code over $\F$. Suppose that $\CC$ is a $\lambda$-constacyclic code and let $x=\varphi(c_{0}, c_{1}, \cdots, c_{n-1})$. Then, $x=c_{0}\overline{1}+c_{1}\overline{g}+\cdots+c_{n-1}\overline{g^{n-1}}$ and 

$\overline{g}\cdot x=$

$=\overline{g}\cdot (c_{0}\overline{1}+c_{1}\overline{g}+\cdots+c_{n-1}\overline{g^{n-1}})=c_{0}\overline{g}\cdot\overline{1}+c_{1}\overline{g}\cdot \overline{g}+\cdots+c_{n-1}\overline{g}\cdot\overline{g^{n-1}}$

$=c_{0}\cdot \overline{g}+c_{1}\overline{g^{2}}+\cdots+c_{n-1}\lambda\cdot \overline{1}$

$=\varphi(\lambda c_{n-1}, c_{0},\cdots, c_{n-2})$. 

Since $\CC$ is $\lambda$-constacyclic, by hypothesis, we have $(\lambda c_{n-1}, c_{0},\cdots, c_{n-2}) \in \CC$, this implies $\varphi(\CC)$ is an ideal of $R^{\gamma}C_{n}$. 

On the other hand, if $(c_{0}, c_{1}, \cdots, c_{n-1}) \in \CC$, then $\overline{g}\cdot \varphi(c_{0}, c_{1}, \cdots, c_{n-1}) \in \varphi(\CC)$, so $(\lambda c_{n-1}, c_{0},\cdots, c_{n-2}) \in \CC$ and $\CC$ is $\lambda$-constacyclic.
\end{proof}

Now, we shall study \textit{isometries} between constacyclic codes.

\begin{definition}
	Let $R$ be a finite commutative ring, $G$ be a finite group and let $\lambda, \, \mu$ be elements of ${\mathcal{U}}(R)$. We say that an isomorphism of $R$ algebra 
		
	\begin{center}
		$\varphi: R^{\lambda} G \longrightarrow R^{\mu} G $. 
	\end{center}
	
	\noindent is an isometry if it preserves the Hamming distance on the algebras, i.e., 
	
	\begin{center}
		$d_{H}(\varphi(a),\varphi(a'))=d_{H}(a,a')$.
	\end{center}
\end{definition}

In \cite{GM}, Ginosar and Moreno have obtained a criterion for isometries between \textit{crossed products}. Since twisted group algebras consist a particular case o crossed product, the result is also true for constacyclic codes. We shall prove that result using only the twisted group algebra structure.

To do that, given a commutative ring $R$ and a group $G$, we say the twisted group algebras $R^{\gamma_{1}}G$, with basis $\overline{G}$, and $R^{\gamma_{2}}G$, with basis $\widetilde{G}$, are \textit{equivalent} if there exists an $R$-algebra isomorphism

\begin{center}
$f: R^{\gamma_{1}}G \longrightarrow R^{\gamma_{2}}G$
\end{center}

\noindent and a mapping $\delta: G \longrightarrow {\mathcal{U}}(R)$ such that $f(\overline{g})=\delta(g)\widetilde{g}$, for all $g\in G$.

\begin{lem} \label{L11} {\rm{\cite [Lemma 1.1]{KAR}}}
Let $R$ be a finite commutative ring and $G$ be a group. The twisted group algebras $R^{\gamma_{1}}G$ and $R^{\gamma_{2}}G$ are equivalent if, and only if, $\gamma_{1}$ and $\gamma_{2}$ are cohomologous. 
\end{lem}

\begin{prop} {\rm{\cite [Theorem 3.5]{GM}}}  \label {P11}
Let $R$ be a commutative ring, $G$ be a finite group of order $n$. There exists an isometry between the twisted group algebras $R^{\gamma_{1}}G$ and $R^{\gamma_{2}}G$ if, and only if, $\gamma_{1}$ and $\gamma_{2}$ are cohomologous.   
\end{prop}

\begin{proof}

Let $\varphi:R^{\gamma_{1}}G \longrightarrow R^{\gamma_{2}}G$ be an isometry, $\{\overline{g}, g \in G\}$ a basis of $R^{\gamma_{1}}G$ and $\{\tilde{g}, g \in G\}$ a basis of $R^{\gamma_{2}}G$. Then, $\varphi(\overline{g})=\delta(g)\tilde{g}$, since the weight of $\overline{g}$ is 1. Thus, by Lemma \ref{L11}, $\gamma_{1}$ and $\gamma_{2}$ are cohomologous.

On the other hand, if $\gamma_{1}$ and $\gamma_{2}$ are cohomologous, again by Lemma \ref{L11}, there exists a $R$-isomorphism $\varphi:R^{\gamma_{1}}G \longrightarrow R^{\gamma_{2}}G$ and a mapping $\delta: G \longrightarrow {\mathcal{U}}(R)$ such that $\varphi(\overline{g})=\delta(g)\tilde{g}$. Thus

$\varphi\left(\displaystyle\sum_{g\in C_{n}}a_{g}\overline{g}\right)=\displaystyle\sum_{g\in C_{n}}a_{g}\varphi(\overline{g})=\displaystyle\sum_{g\in C_{n}}a_{g}\delta(g)\tilde{g}$, which implies that $\varphi$ is an isometry.
\end{proof}

\vr

Now, we have the following

\begin{prop} \label{P22}
	Let $R$ be a commutative ring, $C_{n}=\left\langle g \mid g^{n}=1\right\rangle$ be a cyclic group of order $n$ and $\lambda, \, \beta$ elements of ${\mathcal{U}}(R)$. Then, the twisted group algebras $R^{\gamma_{\lambda}}C_{n}$ and $R^{\gamma_{\beta}}C_{n}$ where
	
	\begin{center}
		$\gamma_{\lambda}(g^{j},g^{k})= \left\{
		\begin{array}{lll}
			\lambda,    & {\rm{if}} \,\, j+k\geq n\\
			1, &  {\rm{if}} \,\, j+k < n.
		\end{array}
		\right.$
		$\gamma_{\beta}(g^{j},g^{k})= \left\{
		\begin{array}{lll}
			\beta,    & {\rm{if}} \,\, j+k\geq n\\
			1, &  {\rm{if}} \,\, j+k < n.
		\end{array}
		\right.$
	\end{center}
	
	\noindent are equivalent if, and only if, there exists a unity a of $R$ such that $\lambda=a^{n}\beta$.
\end{prop}	
	
	\begin{proof}
		Suppose that $R^{\gamma_{\lambda}}C_{n}$ and $R^{\gamma_{\beta}}C_{n}$ are equivalent. By Lemma \ref{L11}, there exist a mapping $\delta: C_{n} \longrightarrow {\mathcal{U}}(R)$ such that $\gamma_{\lambda}(g^{j},g^{k})=\delta(g^{j})\delta(g^{k})\delta(g^{j+k})^{-1}\gamma_{\beta}(g^{j},g^{k})$, for all $0\leq i,\,k \leq n-1$.
		So, $1=\gamma_{\lambda}(1,g)=\delta(1)\delta(g)\delta(g)^{-1}\gamma_{\beta}(1,g)=\delta(1)$. Furthermore, $1=\gamma_{\lambda}(g,g)=\delta(g)\delta(g)\delta(g^{2})^{-1}\gamma_{\beta}(g,g) \Rightarrow \delta(g^{2})=\delta(g)^{2}$. Now, $1=\gamma_{\lambda}(g,g^{2})=\delta(g)\delta(g^{2})\delta(g^{3})^{-1}\gamma_{\beta}(g,g^{2})\Rightarrow \delta(g^{3})=\delta(g)^{3}$. 
		
		Consequently, for all $k<n$, we have 
		
		\begin{center}
			$1=\gamma_{\lambda}(g,g^{k-1})=\delta(g)\delta(g^{k-1})\delta(g^{k})^{-1}\gamma_{\beta}(g,g^{k-1}) \Rightarrow \delta(g^{k})=\delta(g)^{k}$.
		\end{center}
		
		This shows us $\lambda=\gamma_{\lambda}(g,g^{n-1})=\delta(g)\delta(g^{n-1})\delta(1)^{-1}\gamma_{\beta}(g,g^{n-1})=\delta(g)^{n}\beta$ and, taking $a=\delta(g)$, we have $\lambda=a^{n}\beta$.
		
		On the other hand, if $\lambda=a^{n}\beta$, for some $a \in {\mathcal{U}}(R)$, we can define $\delta:C_{n} \longrightarrow {\mathcal{U}}(R)$ by $\delta(g^{i})=a^{i}$ and it is not difficult to see that $\gamma_{\lambda}(g^{j},g^{k})=\delta(g^{j})\delta(g^{k})\delta(g^{i+k})^{-1}\gamma_{\beta}(g^{j},g^{k})$, for all $0\leq i,\,k \leq n-1$.
	\end{proof}

\begin{cor} \label{C11}
	Let $R$ be a commutative ring and let $C_{n}=\left\langle g \mid g^{n}=1\right\rangle$ be a cyclic group of order $n$. Then, the twisted group algebra $R^{\gamma}C_{n}$ where 
	
	\begin{center}
		$\gamma_{\lambda}(g^{j},g^{k})= \left\{
		\begin{array}{lll}
			\lambda,    & {\rm{if}} \,\, j+k\geq n\\
			1, &  {\rm{if}} \,\, j+k < n.
		\end{array}
		\right.$
	\end{center}
	
	\noindent is equivalent to the group algebra $RC_{n}$ if, and only if there exists a unity a of $R$ such that $\lambda=a^{n}$. 
\end{cor}

Notice that if we take $R=\F_{q}$, a finite field with $q$ elements, in Proposition \ref{P22} and Corollary \ref{C11}, we have Theorem 3.2 and Corollary 3.4 obtained in \cite{CFLL}. Also, taking $\lambda=-1$, we obtain Lemma 4.8 of \cite{GM}.

\begin{cor} {\rm{\cite [Corollary 3.5]{CFLL}}}
	Let n be a positive integer such that gcd(n,q-1)=1, $\F_{q}$ be a finite field with q elements and let $C_{n}=\left\langle g \mid g^{n}=1\right\rangle$ be a cyclic group of order $n$. Then, the twisted group algebra $\F_{q}^{\gamma}C_{n}$ where 
	
	\begin{center}
		$\gamma_{\lambda}(g^{j},g^{k})= \left\{
		\begin{array}{lll}
			\lambda,    & {\rm{if}} \,\, j+k\geq n\\
			1, &  {\rm{if}} \,\, j+k < n.
		\end{array}
		\right.$
	\end{center}
	
	\noindent is equivalent to the group algebra $\F_{q}C_{n}$. 
	
\end{cor}

\section{The k-Galois form}

In this section, we shall give definitions and some known results which have elementary proofs in twisted group algebras language.

Let $\F_{q}$ be a finite field with $q=p^{m}$ elements, $G$ be a finite group and $\F_{q}^{\gamma}G$ the twisted group algebra of $G$ over $\F_{q}$. Given $\alpha=\displaystyle\sum_{g \in G}\alpha_{g}\overline{g}$, $\beta=\displaystyle\sum_{g \in G}\beta_{g}\overline{g}$ two elements  of $\F_{q}^{\gamma}G$, for each $k$, $0\leq k< m$, we define the \textit{k-Galois form} on $\F_{q}^{\gamma}G$ as 

\vr

\begin{center}
$[\alpha,\beta]_{k}=\displaystyle\sum_{g \in G}\alpha_{g}\beta_{g}^{p^{k}}$.
\end{center}

\vr

It is not difficult to see that $k$-Galois form is just the Euclidean inner product if $k=0$. Thus, given a twisted group code $\CC$, we can define the \textit{k-Galois dual code of} $\CC$ as

\begin{center}
$\CC^{\perp_{k}}=\{\beta \in \F_{q}^{\gamma}G \mid [\alpha,\beta]_{k}=0, \, \forall \, \alpha \in \CC \}$.
\end{center}

Given two non-zero elements $\lambda$ and $\beta$ of $\F_{q}$, we say that a linear code $\CC$ is $\lambda-\beta$-\textit{constacyclic} if $\CC$ is $\lambda-$constacyclic and $\beta$-constacyclic. Dinh, in \cite{D1}, proved if $\lambda\neq \beta$, the only $\lambda-\beta$-constacyclic codes are $\{0\}$ and $\F_{q}^{n}$.

In terms of twisted group algebras, we have 

\begin{lem} \label{L2} {\rm{\cite [Proposition ]{D1}}}
Let $\F_{q}$ be a finite field with $p=q^{m}$ elements and let $C_{n}=\left\langle g \mid g^{n}=1\right\rangle$ be a cyclic group of order $n$ and $\lambda, \, \beta$ non-zero elements of $\F_{q}$. Consider the twisted group algebras $\F_{q}^{\gamma_{\lambda}}C_{n}$ and $\F_{q}^{\gamma_{\beta}}C_{n}$ where

\vr

\begin{center}
		$\gamma_{\lambda}(g^{j},g^{k})= \left\{
		\begin{array}{lll}
			\lambda,    & {\rm{if}} \,\, j+k\geq n\\
			1, &  {\rm{if}} \,\, j+k < n.
		\end{array}
		\right.$
		$\gamma_{\beta}(g^{j},g^{k})= \left\{
		\begin{array}{lll}
			\beta,    & {\rm{if}} \,\, j+k\geq n\\
			1, &  {\rm{if}} \,\, j+k < n.
		\end{array}
		\right.$
	\end{center}
	
\vr	
	
If $\CC$ is a non-zero $\lambda$-constacyclic and also $\beta$-constacyclic code, then $\lambda=\beta$.	
\end{lem}

\begin{proof}
Let $c=\displaystyle\sum_{i=0}^{n-1}c_{i}\overline{g}^{i}$ a non-zero element of $\CC$. Since, by hypothesis, $\CC$ is $\lambda$-constacyclic and also $\beta$-constacyclic code, we have that

\vr

$\overline{g}\cdot c= c_{0}\overline{g}\cdot\overline{1}+c_{1}\overline{g}\cdot \overline{g}+\cdots+c_{n-1}\overline{g}\cdot\overline{g^{n-1}}$

$\hspace{0.7cm}=c_{0}\cdot \overline{g}+c_{1}\overline{g^{2}}+\cdots+c_{n-1}\lambda\cdot \overline{1}$

$\hspace{0.7cm}=c_{0}\cdot \overline{g}+c_{1}\overline{g^{2}}+\cdots+c_{n-1}\beta\cdot \overline{1}$

so, $\lambda=\beta$ since the set $\{\overline{g}, g \in C_{n}\}$ is a basis of $\F_{q}^{\gamma_{\lambda}}C_{n}$ and $\F_{q}^{\gamma_{\beta}}C_{n}$.
\end{proof}

\begin{prop} \label{P1} {\rm{\cite [Lemma 4.3 ]{FZ}}}
Let $\F_{q}$ be a finite field with $q=p^{m}$ elements, $C_{n}=\left\langle g, \, g^{n}=1\right\rangle$ be a cyclic group of order n and $\F_{q}^{\gamma_{\lambda}}C_{n}$ the twisted group algebra of $C_{n}$ over $\F_{q}$ where

\begin{center}
		$\gamma_{\lambda}(g^{j},g^{k})= \left\{
		\begin{array}{lll}
			\lambda,    & {\rm{if}} \,\, j+k\geq n\\
			1, &  {\rm{if}} \,\, j+k < n.
		\end{array}
		\right.$
\end{center}		

Then, if $\CC$ is a $\lambda$-constacyclic code, its k-Galois dual $\CC^{\perp_{k}}$ is a $\lambda^{-p^{m-k}}$-constacyclic code.
\end{prop}

\begin{proof}
Let $\F_{q}^{\gamma_{\lambda^{-p^{m-k}}}}C_{n}$ be the twisted group algebra with twisting defined by

\begin{center}
		$\gamma_{\lambda^{-p^{m-k}}}(g^{j},g^{k})= \left\{
		\begin{array}{lll}
			\lambda^{-p^{m-k}},    & {\rm{if}} \,\, j+k\geq n\\
			1, &  {\rm{if}} \,\, j+k < n.
		\end{array}
		\right.$
\end{center}

Let $\CC$ be a $\lambda$-constacyclic code and, given any element $c=\displaystyle\sum_{i=0}^{n-1}c_{i}\overline{g}^{i}$ of the code $\CC$ and $x=\displaystyle\sum_{i=0}^{n-1}x_{i}\overline{g}^{i}$ an element of $\CC^{\perp_{k}}$ , we have
\vr

$[c,\overline{g}x]_{k}=c_{0}x_{n-1}^{p^{k}}(\lambda^{-p^{m-k}})^{p^{k}}+c_{1}x_{0}^{p^{k}}+\cdots+c_{n-1}x_{n-2}^{p^{k}}$

$\hspace{1.15cm}=c_{0}x_{n-1}^{p^{k}}\lambda^{-p^{m}}+c_{1}x_{0}^{p^{k}}+\cdots+c_{n-1}x_{n-2}^{p^{k}}$

$\hspace{1.15cm}=c_{1}x_{0}^{p^{k}}+c_{2}x_{1}^{p^{k}}+\cdots+\lambda^{-1}c_{0}x_{n-1}^{p^{k}}$

$\hspace{1.15cm}=[\overline{g}^{-1}c,x]_{k}=0$.

\vr

Then, $\overline{g}x \in \CC^{\perp_{k}}$ and the proof is completed.
\end{proof}

\begin{definition}
Let $\CC$ be a constacyclic code over a finite field $\F_{q}$. We say that $\CC$ is a linear complementary k-Galois dual code (k-Galois LCD code for shorty) if $\CC \cap \CC^{\perp_{k}}=\{0\}$.
\end{definition}

By Lemma \ref{L2} and Proposition \ref{P1}, we get

\begin{cor} {\rm{\cite [Corollary 3.3]{LFL}}} \label{COR2}
If $\lambda^{1+p^{m-k}}\neq 1$, then any $\lambda$-constacyclic code $\CC$ over $\F_{q}$ is a k-Galois LCD code.
\end{cor}

Notice that, since $\CC^{\perp_{k}}$ is a linear subspace and the $k-$Galois form is non-degenerate, we have that dim $\CC$ + dim $\CC^{\perp_{k}}=n$.

\section{The classical involution}

Let $R$ be a commutative ring with identity and let $G$ be a group. Consider the following mapping $^{\ast} : R^{\gamma}G \longrightarrow R^{\gamma}G $ given by $\left(\displaystyle\sum_{g \in G} \alpha_{g}\overline{g}\right)^{\ast}=\displaystyle\sum_{g \in G} \alpha_{g}\overline{g}^{-1}$. 

\vr

It is not difficult to see the mapping $^\ast$ above defined has the following property  $(\alpha+\beta)^{\ast}=\alpha^{\ast}+\beta^{\ast}$

\begin{enumerate}
	\item [(i)] $(\alpha+\beta)^{\ast}=\alpha^{\ast}+\beta^{\ast}$;
\end{enumerate}

Now, since, by Lemma \ref{1.1}, $\overline{g}^{-1}=\gamma(g,g^{-1})^{-1}\overline{g^{-1}}=\gamma(g^{-1},g)^{-1}\overline{g^{-1}}$, we have that

$\left(\overline{g}^{\ast}\right)^{\ast}=\left(\overline{g}^{-1}\right)^{\ast}=\left(\gamma(g,g^{-1})^{-1}\overline{g^{-1}}\right)^{\ast}$


\hspace{1cm}$=\gamma(g^{-1},g)^{-1}\overline{g^{-1}}^{-1}$

\hspace{1cm}$=\gamma(g^{-1},g)^{-1}\gamma(g^{-1},g)^{-1}\overline{(g^{-1})^{-1}}=\gamma(g,g^{-1})^{-2}\overline{g}$

\noindent for all $g \in G$. So, we can conclude if $\gamma(g,g^{-1})^{2}=1$, then $(\alpha^{\ast})^{\ast}=\alpha$, for all $\alpha \in R^{\gamma}G$. Now, 

$\left(\overline{g}\cdot \overline{h}\right)^{\ast}=\left(\gamma(g,h)\overline{gh}\right)^{\ast}=\gamma(g,h)\overline{gh}^{-1}$


\hspace{1.35cm}$=\gamma(g,h)\gamma(gh,h^{-1}g^{-1})\overline{h^{-1}g^{-1}}$


Since $\gamma$ is a 2-cocycle, we get that

\begin{center}
$\gamma(gh,h^{-1}g^{-1})=\gamma(g,g^{-1})\gamma(h,h^{-1})\gamma(g,h)^{-1}\gamma(h^{-1},g^{-1})^{-1}$,
\end{center}

\noindent so $\left(\overline{g}\cdot \overline{h}\right)^{\ast}=\gamma(g,g^{-1})\gamma(h,h^{-1})\gamma(h^{-1},g^{-1})^{-1}\overline{h^{-1}g^{-1}}$.

\vr

On the other hand, 

\vr

$\overline{h}^{\ast}\overline{g}^{\ast}=\overline{h}^{-1}\cdot \overline{g}^{-1}=\gamma(h,h^{-1})\gamma(g,g^{-1})\overline{h^{-1}}\cdot \overline{g^{-1}}$


\hspace{1cm}$=\gamma(h,h^{-1})\gamma(g,g^{-1})\gamma(h^{-1},g^{-1})\overline{h^{-1}g^{-1}}$.

\vr

Thus, $\left(\overline{g}\cdot \overline{h}\right)^{\ast}=\overline{h}^{\ast}\overline{g}^{\ast}$ if and only if $\gamma(h^{-1},g^{-1})^{-1}=\gamma(h^{-1},g^{-1})$, for all $g$, $h \in G$. Consequently, we conclude $(\alpha\beta)^{\ast}=\beta^{\ast}\alpha^{\ast}$, for all $\alpha, \, \beta \in R^{\gamma}G$ if, and only if, $\gamma(g,h)^{2}=1$, for all $g$, $h \in G$.

\begin{definition}
Let R be a commutative ring with identity and let G be a group. The mapping $^{\ast} : R^{\gamma}G \longrightarrow R^{\gamma}G $ given by $\left(\displaystyle\sum_{g \in G} \alpha_{g}\overline{g}\right)^{\ast}=\displaystyle\sum_{g \in G} \alpha_{g}\overline{g}^{-1}$ with $\gamma(g,h)^{2}=1$, is called the classical involution of $R^{\gamma}G$.
\end{definition}

\begin{lem} \label{L1}
Let $\F_{q}$ be a finite field with $q=p^{m}$ elements, $C_{n}=\left\langle g, \, g^{n}=1\right\rangle$ be a cyclic group of order n and $\F_{q}^{\gamma_{\lambda}}C_{n}$ the twisted group algebra of $C_{n}$ over $\F_{q}$ where

\begin{center}
		$\gamma_{\lambda}(g^{j},g^{k})= \left\{
		\begin{array}{lll}
			\lambda,    & {\rm{if}} \,\, j+k\geq n\\
			1, &  {\rm{if}} \,\, j+k < n.
		\end{array}
		\right.$
\end{center}		

Given two arbitrary elements $\alpha=\displaystyle\sum_{i=0}^{n-1}\alpha_{i}\overline{g}^{i}$ and $\beta=\displaystyle\sum_{i=0}^{n-1}\beta_{i}\overline{g}^{i}$ of $\F_{q}^{\gamma_{\lambda}}C_{n}$, let us denote by $\beta^{(p^{k})}$ the element $\displaystyle\sum_{i=0}^{n-1}\beta_{i}^{p^{k}}\overline{g}^{i}$. If $\alpha\left(\beta^{(p^{k})}\right)^{\ast}=0$ and $\lambda^{2}=1$, then $[\alpha,\beta]_{k}=0$.
\end{lem}

\begin{proof}
It is not difficult to see, the coefficient of $1=\overline{1}$ in the product $\alpha\left(\beta^{(p^{k})}\right)^{\ast}$ is exactly $[\alpha,\beta]_{k}$. Since, by hypothesis, $\alpha\left(\beta^{(p^{k})}\right)^{\ast}=0$, we have $[\alpha,\beta]_{k}=0$. 
\end{proof}

\section{Euclidean Constacyclic LCD codes.}

In this section, we shall characterize constacyclic LCD codes in terms of its idempotent generator with respect to Euclidean inner product.

Let $\F_{q}$ be a finite field, $C_{n}=\left\langle g, \, g^{n}=1\right\rangle$ be a cyclic group of order $n$ and $\F_{q}^{\gamma_{\lambda}}C_{n}$ the twisted group algebra of $C_{n}$ over $\F_{q}$ where

\begin{center}
		$\gamma_{\lambda}(g^{j},g^{k})= \left\{
		\begin{array}{lll}
			\lambda,    & {\rm{if}} \,\, j+k\geq n\\
			1, &  {\rm{if}} \,\, j+k < n.
		\end{array}
		\right.$
\end{center}		

\noindent for some non-zero $\lambda \in \F_{q}$. Given $\alpha=\displaystyle\sum_{g \in C_{n}}\alpha_{g}\overline{g}$, $\beta=\displaystyle\sum_{g \in C_{n}}\beta_{g}\overline{g}$ two elements  of $\F_{q}^{\gamma_{\lambda}}C_{n}$, we define the \textit{Euclidean inner product} on $\F_{q}^{\gamma_{\lambda}}C_{n}$ as 

\vr

\begin{center}
$[\alpha,\beta]=\displaystyle\sum_{g \in G}\alpha_{g}\beta_{g}$.
\end{center}

Let $\CC$ be a constacyclic code over $\F_{q}^{\gamma}C_{n}$, that is, an ideal of $\F_{q}^{\gamma_{\lambda}}C_{n}$ . It is well-known that the set $\CC^{\perp}=\{x \in R^{\gamma}G \mid [x,\alpha]=0, \forall \alpha \in \CC\}$ is an ideal in the twisted group algebra $\F_{q}^{\gamma_{\lambda^{-1}}}C_{n}$ where

\begin{center}
		$\gamma_{\lambda^{-1}}(g^{j},g^{k})= \left\{
		\begin{array}{lll}
			\lambda^{-1},    & {\rm{if}} \,\, j+k\geq n\\
			1, &  {\rm{if}} \,\, j+k < n.
		\end{array}
		\right.$
\end{center}		

\begin{definition}
Let $\CC$ be a constacyclic code over a finite field $\F_{q}$. We say that $\CC$ is a linear complementary dual code (LCD code for shorty) if $\CC \cap \CC^{\perp}=\{0\}$.
\end{definition}

Notice that, the Corollary \ref{COR2} shows us if $\lambda^{2}\neq 1$, any $\lambda$-constacyclic code $\CC$ is LCD.

\begin{prop} \label{LCDE}
Let $\F_{q}$ be a finite field, $C_{n}=\left\langle g, \, g^{n}=1\right\rangle$ be a cyclic group of order n and $\F_{q}^{\gamma_{\lambda}}C_{n}$ the twisted group algebra of $C_{n}$ over $\F_{q}$ where

\begin{center}
		$\gamma_{\lambda}(g^{j},g^{k})= \left\{
		\begin{array}{lll}
			\lambda,    & {\rm{if}} \,\, j+k\geq n\\
			1, &  {\rm{if}} \,\, j+k < n.
		\end{array}
		\right.$
\end{center}		

\noindent for some non-zero $\lambda \in \F_{q}$. If $\lambda^{2}=1$, then $\CC$ is a $\lambda$-constacyclic LCD code if, and only if, $\CC$ is generated by an idempontent e such that $e=e^{\ast}$.
\end{prop}

\begin{proof}
First of all, if $\CC$ is a $\lambda$-constacyclic code which is also LCD, we have the following decomposition of ideals $\F_{q}^{\gamma_{\lambda}}C_{n}=\CC \oplus \CC^{\perp}$ since $\lambda^{2}=1$. So , it is well-know, there exist idempotents $e$ and $f$ such that $1=e+f$, $e\cdot f=0$, $\CC=\left\langle e\right\rangle$ and $\CC^{\perp}=\left\langle f\right\rangle$. 

Since $[e,1-e]=0$, we have that $[1,e^{\ast}(1-e)]=0$. Now, given $a=\displaystyle\sum_{i=0}^{n-1}a_{i}\overline{g}^{i}$ and $b=\displaystyle\sum_{i=0}^{n-1}b_{i}\overline{g}^{i}$ two arbitrary elements of $\F_{q}^{\gamma_{\lambda}}C_{n}$, then

\begin{center}
$[\overline{g}a,\overline{g}b]_{k}=a_{0}b_{0}+a_{1}b_{1}+\cdots+a_{n-2}b_{n-2}+(a_{n-1}\lambda)(b_{n-1}\lambda)=[a,b]$
\end{center} 
 
So $[\overline{g},\overline{g}e^{\ast}(1-e)]=0$, for all $g \in G$. Since the Euclidean inner product is non-degenerated, we get that $e^{\ast}(1-e)=0$ and $e^{\ast}=e^{\ast}e$, which implies that  $e^{\ast}=(ee^{\ast})^{\ast}=ee^{\ast}=e$.

On the other hand, if $e$ is an idempotent such that $e=e^{\ast}$ and $\CC=\left\langle e\right\rangle$, then, $e(1-e)^{\ast}=e(1-e^{\ast})=e(1-e)=0$ and, by Lemma \ref{L1}, $[e,1-e]=0$. Writing $1=e+(1-e)$ we have $\F_{q}^{\gamma_{\lambda}}C_{n}=\CC \oplus \F_{q}^{\gamma_{\lambda}}C_{n}(1-e)$. Since $[e,1-e]=0$ and  $dim_{\F_{q}}\CC + dim_{\F_{q}}\F_{q}^{\gamma_{\lambda}}C_{n}(1-e)=n$, we conclude that $\CC^{\perp}=\F_{q}^{\gamma_{\lambda}}C_{n}(1-e)$.

Thus, $\CC$ is a $\lambda$-constacyclic LCD code. 
\end{proof}

\begin{cor} \label{C1} {\rm{\cite [Theorem 3.1 ]{CW}}}
Let $\F_{q}$ be a finite field with $q=p^{m}$ elements, $C_{n}=\left\langle g, \, g^{n}=1\right\rangle$ be a cyclic group of order n. A cyclic code $\CC$ is a LCD code with respect the Euclidean inner product, if, and only if,  $\CC=\left\langle e \right\rangle$ such that $e^{2}=e$ and $e=e^{\ast}$.
\end{cor}

\section{Some good LCD codes}

In this section, we shall exhibit some good LCD codes obtained from twisted group algebras.

\vr

\noindent \textit{Example 1:} Let $C_{10}=\left\langle g, \,\, g^{10}=1\right\rangle$ be a cyclic group of order 10 and let $\F_{3}$ be a finite field with 3 elements. Consider the twisted group algebra $\F_{3}^{\gamma_{2}}C_{10}$, thus in this case, we have $\overline{g}^{10}=2$. Finally, taking the elements $e=\overline{g}^{8}+2\overline{g}^{6}+\overline{g}^{4}+2\overline{g}^2+2$ and $f=2\overline{g}^8 + \overline{g}^6 + 2\overline{g}^4 + \overline{g}^2 + 2 $.

It is not difficult to see $e^{2}=e$ and 

\begin{center}
$e^{\ast}=2\cdot 2\overline{g}^{2}+2\cdot\overline{g}^{4}+2\cdot 2\cdot \overline{g}^{6}+ 2\cdot \overline{g}^{8}+ 2= e$.
\end{center}

So, by Proposition \ref{LCDE}, the code $\CC$ generated by $e$ is a LCD code of dimension of dimension 8 and weight 2 which are exactly the parameters of the best [10,8] code known.

Finally, notice that $f=1-e$, so it is also an idempotent with $f^{\ast}=f$ and the code generated by $f$ is LCD of dimension 2 and weight 5 and the best [10,2] code known has weight 7.

\vr

\noindent \textit{Example 2:} Let $C_{9}=\left\langle g, \,\, g^{9}=1\right\rangle$ be a cyclic group of order 5 and let $\F_{5}$ be a finite field with 5 elements. Consider the twisted group algebra $\F_{5}^{\gamma_{4}}C_{9}$, thus in this case, we have $\overline{g}^{9}=4$. Finally, taking the elements $e=\overline{g}^8 + 4\overline{g}^7 + 3\overline{g}^6 + 4\overline{g}^5 + \overline{g}^4 + 2\overline{g}^3 + \overline{g}^2 + 4\overline{g} + 3$ and $f=4\overline{g}^8 + \overline{g}^7 + 2\overline{g}^6 + \overline{g}^5 + 4\overline{g}^4 + 3\overline{g}^3 + 4\overline{g}^2 + \overline{g} + 3$.

It is not difficult to see $e^{2}=e$ and 

\begin{center}
$e^{\ast}= 4\overline{g}+4\cdot 4\cdot\overline{g}^{2}+3\cdot 4\cdot \overline{g}^{3}+ 4\cdot 4\cdot \overline{g}^{4}+ 4\cdot \overline{g}^{5} + 2\cdot 4\cdot \overline{g}^{6} + 4\cdot \overline{g}^{7} + 4\cdot 4\cdot \overline{g}^{8}+ 3= e$.
\end{center}

So, by Proposition \ref{LCDE}, the code $\CC$ generated by $e$ is a LCD code of dimension of dimension 7 and weight 2 which are exactly the parameters of the best [9,7] code known.

Finally, notice that $f=1-e$, so it is also an idempotent with $f^{\ast}=f$ and the code generated by $f$ is LCD of dimension 2 and weight 6 and the best [6,2] code known has weight 7.

\vr

\noindent \textit{Example 3:} Let $C_{21}=\left\langle g, \,\, g^{21}=1\right\rangle$ be a cyclic group of order 21 and let $\F_{5}$ be a finite field with 5 elements. Consider the twisted group algebra $\F_{5}^{\gamma_{4}}C_{21}$, thus in this case, we have $\overline{g}^{21}=4$. Finally, taking the element 

\vr

\noindent $e=4\overline{g}^{19} + 4\overline{g}^{18} + \overline{g}^{15} + 2\overline{g}^{14} + 4\overline{g}^{13} + 4\overline{g}^{12} + 4\overline{g}^{11} + \overline{g}^{10} + \overline{g}^9 + \overline{g}^8 + 3\overline{g}^7 + 4\overline{g}^6 + \overline{g}^3 + \overline{g}^2 + 1 $ 

\vr

\noindent and 

\vr

\noindent $f=\overline{g}^{19} + \overline{g}^{18} + 4\overline{g}^{15} + 3\overline{g}^{14} + \overline{g}^{13} + \overline{g}^{12} + \overline{g}^{11} + 4\overline{g}^{10} + 4\overline{g}^9 + 4\overline{g}^8 + 2\overline{g}^7 + \overline{g}^6 + 4\overline{g}^3 + 4\overline{g}^2 $ .

It is not difficult to see $e^{2}=e$ and 

\vr

\noindent $e^{\ast}= 4\cdot 4\overline{g}^{2}+4\cdot 4\cdot\overline{g}^{3}+ 4\cdot \overline{g}^{6}+ 2\cdot 4\cdot \overline{g}^{7}+ 4\cdot \cdot 4 \overline{g}^{8} + 4\cdot 4\cdot \overline{g}^{9} + 4\cdot \cdot 4 \overline{g}^{10} + 4\cdot \overline{g}^{11}$ 

$ + 4\cdot \overline{g}^{12}+ 4\cdot \overline{g}^{13}+ 3\cdot 4\cdot \overline{g}^{14}+ 4\cdot 4\cdot \overline{g}^{15} + 4\cdot \overline{g}^{18} + 4\cdot \overline{g}^{19}= e$.

So, by Proposition \ref{LCDE}, the code $\CC$ generated by $e$ is a LCD code of dimension of dimension 6 and weight 12 which are exactly the parameters of the best [21,6] code known.

Finally, notice that $f=1-e$, so it is also an idempotent with $f^{\ast}=f$ and the code generated by $f$ is LCD of dimension 15 and weight 3 and the best [21,15] code known has weight 5.

\vr

\noindent \textit{Example 4:} Let $C_{19}=\left\langle g, \,\, g^{19}=1\right\rangle$ be a cyclic group of order 19 and let $\F_{7}$ be a finite field with 7 elements. Consider the twisted group algebra $\F_{7}^{\gamma_{6}}C_{19}$, thus in this case, we have $\overline{g}^{19}=6$. Finally, taking the elements

\newpage

\noindent $e=3\overline{g}^{18} + 6\overline{g}^{17} + \overline{g}^{16} + 5\overline{g}^{15} + \overline{g}^{14} + 5\overline{g}^{13} + 3\overline{g}^{12} + 4\overline{g}^{11} + 2\overline{g}^{10} + 5\overline{g}^9 + 3\overline{g}^8 + 4\overline{g}^7$

$ + 2\overline{g}^6 + 6\overline{g}^5 + 2\overline{g}^4 + 6\overline{g}^3 + \overline{g}^2 + 4\overline{g}   $ 

\vr

\noindent and 

\vr

\noindent $f=4\overline{g}^{18} + \overline{g}^{17} + 6\overline{g}^{16} + 2\overline{g}^{15} + 6\overline{g}^{14} + 2\overline{g}^{13} + 4\overline{g}^{12} + 3\overline{g}^{11} + 5\overline{g}^{10} + 2\overline{g}^9 + 4\overline{g}^8 + 3\overline{g}^7$

$ + 5\overline{g}^6 + \overline{g}^5 + 5\overline{g}^4 + \overline{g}^3 + 6\overline{g}^2 + 3\overline{g} + 1  $ .

It is not difficult to see $e^{2}=e$ and 

\vr

\noindent $e^{\ast}= 3\cdot 6\overline{g}+6\cdot 6\overline{g}^{2}+ 6 \overline{g}^{3}+ 5\cdot 6 \overline{g}^{4}+ 6\overline{g}^{5} + 5\cdot 6\overline{g}^{6} + 3\cdot 6 \overline{g}^{7} + 4\cdot 6 \overline{g}^{8}$

$ + 2\cdot 6 \overline{g}^{9} + 5\cdot 6 \overline{g}^{10} + 3\cdot 6 \overline{g}^{11} + 4\cdot 6 \overline{g}^{12} + 2\cdot 6 \overline{g}^{13} + 6\cdot 6 \overline{g}^{14} + 2\cdot 6 \overline{g}^{15} + 6\cdot 6 \overline{g}^{16}$ 

$ + 6\cdot \overline{g}^{17}+ 4\cdot 6 \overline{g}^{18}= e$.

So, by Proposition \ref{LCDE}, the code $\CC$ generated by $e$ is a LCD code of dimension of dimension 7 and weight 10 which are exactly the parameters of the best [19,7] code known.

Finally, notice that $f=1-e$, so it is also an idempotent with $f^{\ast}=f$ and the code generated by $f$ is LCD of dimension 12 and weight 6 which are exactly the parameters of the best [19,12] code known.

\section{k-Galois constacyclic LCD codes}

In this section, we shall prove some results about $k$-Galois constacyclic LCD codes.

\begin{thrm} \label{T1}
Let $\F_{q}$ be a finite field with $q=p^{m}$ elements, $C_{n}=\left\langle g, \, g^{n}=1\right\rangle$ be a cyclic group of order n and $\F_{q}^{\gamma_{\lambda}}C_{n}$ the twisted group algebra of $C_{n}$ over $\F_{q}$ where

\begin{center}
		$\gamma_{\lambda}(g^{j},g^{k})= \left\{
		\begin{array}{lll}
			\lambda,    & {\rm{if}} \,\, j+k\geq n\\
			1, &  {\rm{if}} \,\, j+k < n.
		\end{array}
		\right.$
\end{center}		

Let e be an idempotent of $\F_{q}^{\gamma_{\lambda}}C_{n}$ and $\lambda^{2}=1$. Then $e=e(e^{(p^{k})})^{\ast}$ if, and only if $[e,1-e]_{k}=0$. 
\end{thrm}

\begin{proof}
Suppose that $e$ is an idempotent such that $e=e(e^{(p^{k})})^{\ast}$. Then,\\ $e-e(e^{(p^{k})})^{\ast}=e(1-(e^{(p^{k})})^{\ast})=e\left((1-e)^{(p^{k})}\right)^{\ast}=0$ and, by Lemma \ref{L1}, $[e,1-e]_{k}=0$.

On the other hand, if $[e,1-e]_{k}=0$, we have that $[1,e^{\ast}(1-e)^{(p^{k})}]_{k}=0$. Now, given $a=\displaystyle\sum_{i=0}^{n-1}a_{i}\overline{g}^{i}$ and $b=\displaystyle\sum_{i=0}^{n-1}b_{i}\overline{g}^{i}$ two arbitrary elements of $\F_{q}^{\gamma_{\lambda}}C_{n}$, since $\lambda^{p^{k}}=\lambda^{-1}$, then 

$[\overline{g}a,\overline{g}b]_{k}=a_{0}b_{0}^{p^{k}}+a_{1}b_{1}^{p^{k}}+\cdots+a_{n-2}b_{n-2}^{p^{k}}+(a_{n-1}\lambda)(b_{n-1}^{p^{k}}\lambda^{p^{k}})=[a,b]_{k}$
 
So $[\overline{g},\overline{g}e^{\ast}(1-e)^{(p^{k})}]_{k}=0$, for all $g \in G$. Since the $k$-Galois form is non-degenerated, we get that $e^{\ast}(1-e)^{(p^{k})}=0$ and $e^{\ast}=e^{\ast}e^{(p^{k})}$. Then, $e=(e^{\ast})^{\ast}=(e^{\ast}e^{(p^{k})})^{\ast}=e(e^{(p^{k})})^{\ast}$. 
\end{proof}





Now, we have the following

\begin{prop}
Let $\F_{q}$ be a finite field with $q=p^{m}$ elements, $C_{n}=\left\langle g, \, g^{n}=1\right\rangle$ be a cyclic group of order n and $\F_{q}^{\gamma_{\lambda}}C_{n}$ the twisted group algebra of $C_{n}$ over $\F_{q}$ where

\begin{center}
		$\gamma_{\lambda}(g^{j},g^{k})= \left\{
		\begin{array}{lll}
			\lambda,    & {\rm{if}} \,\, j+k\geq n\\
			1, &  {\rm{if}} \,\, j+k < n.
		\end{array}
		\right.$
\end{center}		

If $\lambda^{2}=1$, then $\CC$ is a $\lambda$-constacyclic code  generated by an idempontent e such that $e=e(e^{(p^{k})})^{\ast}$ if, and only if $\CC$ is k-Galois LCD code.
\end{prop}

\begin{proof}
First of all, if $\CC$ is a $\lambda$-constacyclic code which is also $k$-Galois LCD, we have the following decomposition of ideals $\F_{q}^{\gamma}C_{n}=\CC \oplus \CC^{\perp_{k}}$ since $\lambda^{1+p^{m-k}}=1$.

It is well-know that there exist idempotents $e$ and $f$ such that $1=e+f$, $e\cdot f=0$, $\CC=\left\langle e\right\rangle$ and $\CC^{\perp_{k}}=\left\langle f\right\rangle$. Then, writing $f=1-e$, we get $[e,1-e]_{k}=0$, so, by Theorem \ref{T1}, the equality $e=e(e^{(p^{k})})^{\ast}$ holds. 

If $\CC$ is generated by an idempontent $e$ such that $e=e(e^{(p^{k})})^{\ast}$, by Theorem \ref{T1}, we have $[e,1-e]_{k}=0$. Then, writing $1=e+(1-e)$ we have $\F_{q}^{\gamma}C_{n}=\CC \oplus \F_{q}^{\gamma}C_{n}(1-e)$. Since $[e,1-e]_{k}=0$ and  $dim_{\F_{q}}\CC + dim_{\F_{q}}\F_{q}^{\gamma}C_{n}(1-e)=n$, we conclude that $\CC^{\perp_{k}}=\F_{q}^{\gamma}C_{n}(1-e)$. Thus, $\CC$ is a $\lambda$-constacyclic $k$-Galois LCD code. 
\end{proof}

\section*{Data availability.}

All data are available from the authors upon reasonable request.

\section*{Conflict of interest}

All authors have participated in (a) conception and design, or analysis and interpretation of the data; (b) drafting the article or revising it critically for important intellectual content; and (c) approval of the final version.  

This manuscript has not been submitted to, nor is under review at, another journal or other publishing venue.

The authors have no affiliation with any organization with a direct or indirect financial interest in the subject matter discussed in the manuscript

\end{document}